\documentclass[english,prl,showpacs,reprint,superscriptaddress]{revtex4-2}
\usepackage[latin9]{inputenc}
\usepackage{color}
\usepackage{babel, array, multirow}
\usepackage{verbatim}
\usepackage{bm}
\usepackage{amsmath}
\usepackage{amssymb}
\usepackage{graphicx}
\usepackage{gensymb}
\usepackage[unicode=true,pdfusetitle,bookmarks=true,bookmarksnumbered=false,bookmarksopen=false,breaklinks=false,pdfborder={0 0 0},pdfborderstyle={},backref=false,colorlinks=true]{hyperref}
\hypersetup{citecolor=blue}
\makeatletter
\usepackage{epstopdf}
\usepackage{amsfonts}
\usepackage{mathrsfs}\usepackage{bm}\usepackage{appendix}\usepackage{color}
\usepackage{txfonts}
\usepackage{float}
\usepackage{epstopdf}\usepackage{ulem}
\usepackage{textcomp}

\newcommand{\Rmnum}[1]{\expandafter\@slowromancap\romannumeral #1@}
\usepackage{CJK}
\usepackage{colortbl}
\makeatother

\begin{document}

\title{Axion Quasibound States in the Continuum and Their Enhanced Axion Nonlinearity}

\author{Chang-Yin Ji}
\affiliation{State Key Laboratory of Chips and Systems for Advanced Light Field Display, Key Lab of advanced optoelectronic quantum architecture and measurement (Ministry of Education) and School of Physics, Beijing Institute of Technology, Beijing 100081, China}

\author{Chong Wang}
\affiliation{State Key Laboratory of Chips and Systems for Advanced Light Field Display, Key Lab of advanced optoelectronic quantum architecture and measurement (Ministry of Education) and School of Physics, Beijing Institute of Technology, Beijing 100081, China}

\author{Jiafang Li}
\email{jiafangli@bit.edu.cn}
\affiliation{State Key Laboratory of Chips and Systems for Advanced Light Field Display, Key Lab of advanced optoelectronic quantum architecture and measurement (Ministry of Education) and School of Physics, Beijing Institute of Technology, Beijing 100081, China}
\affiliation{School of Optics and Photonics, Beijing Institute of Technology, Beijing 10008l, China} 

\author{Yugui Yao}
\email{ygyao@bit.edu.cn}
\affiliation{State Key Laboratory of Chips and Systems for Advanced Light Field Display, Key Lab of advanced optoelectronic quantum architecture and measurement (Ministry of Education) and School of Physics, Beijing Institute of Technology, Beijing 100081, China}

\begin{abstract}
Dynamical axion quasiparticles (DAQs) in condensed-matter systmes have emerged as a fascinating platform for exploring axion physics and nonlinear optical phenomena. However, the amplitude of the coherent oscillating $\theta$ field is even smaller than that of the static counterpart, making the experimental observation of DAQs particularly challenging. Here, we introduce symmetry-sensitive bound states in the continuum (BICs) physics to enable the all-optical detection of DAQs and enhance DAQ-mediated second-harmonic generation (SHG). Specifically, we uncover that DAQs induce optical shear that breaks the $C_{2z}$ symmetry of metagratings without geometric engineering, thereby converting symmetry-protected BICs into axion quasi-BICs (qBICs). In this context, the evolution of the quality factor and topological charge in momentum space, together with the appearance of optical resonances in far field associated with axion qBICs, offers achiral optical signatures for DAQ detection beyond the widely used Kerr and Faraday effects. Furthermore, by exploiting axion qBICs in the critical coupling regime to enhance fundamental optical field, we show that the SHG efficiency can be enhanced more than $10^5$-fold compared to bulk axion materials, achieving a conversion efficiency exceeding $10^{-6}~W^{-1}$. The SHG is governed by the magnetoelectric coupling, which differs from that in conventional SHG materials. Our findings establish a new approach for probing DAQs via achiral optical effects, while providing a versatile route for reconfigurable photonics and terahertz nonlinear optics.
\end{abstract}

\maketitle
Axions are spinless bosons arising from the spontaneous breaking of Peccei-Quinn symmetry, originally proposed to solve the strong charge-parity problem in quantum chromodynamics (QCD) \cite{peccei1977cp,peccei1977constraints,wilczek1978problem,weinberg1978new}, and are now of central importance in dark matter and string theory \cite{svrcek2006axions,du2018search,duan2023resonant}. However, the axion field $\theta$ is expected to be vanishingly small in QCD and cosmology. Topological field theory predicts that an effectively large $\theta$ can emerge in the electrodynamics of certain topological materials in condensed matter \cite{qi2008topological}, including three-dimensional topological insulators (TIs), magnetic and antiferromagnetic TIs, and Weyl semimetals \cite{essin2009magnetoelectric,tse2010giant,maciejko2010topological,swingle2011correlated,valdes2012terahertz,shuvaev2013terahertz,morimoto2015topological,wu2016quantized,wang2016dynamical,mogi2017magnetic,grauer2017scaling,dziom2017observation,zhang2019topological,xu2019higher,gooth2019axionic,zhang2020large,wang2020dynamical,gao2021layer,qiu2023axion,shyta2025axion,hu2026half}. $\theta$ is quantized to $\pi$ (mod $2\pi$) in TIs that preserve either time-reversal $\cal{T}$ or space-inversion $\cal{P}$ symmetry \cite{nenno2020axion,sekine2021axion}. A dynamical axion field $\theta(\bm r, t)$ can emerge from coherent collective magnetic excitations when both $\cal{T}$ and $\cal{P}$ symmetries are broken, giving rise to dynamical axion quasiparticles (DAQs) that are exactly analogous to high-energy axion particles \cite{li2010dynamical,qiu2025observation}. $\theta(\bm r, t)$ enables many fascinating phenomena beyond those accessible in the static case, including instability, axion polaritons, ultrafast spin polarization, nonlinear Raman processes and stimulated axion scattering \cite{ooguri2012instability, li2010dynamical, terccas2018axion,liebman2025multiphoton,smith2026theory}. A variety of schemes and phenomena have been proposed to probe DAQs, including X-ray and Gamma ray, Faraday and Kerr rotations, nonlinear responses, microcantilever torque magnetometry, ferroaxionic force and so on \cite{zioutas2005first,fairbairn2007transparency,akerib2017first,du2018search,li2010dynamical,sikivie2021invisible,liebman2025multiphoton,smith2026theory,gao2025detecting,boyanovsky2025probing,arvanitaki2026detecting,hu2026half}. However, recent experiments studies have revealed that the amplitude of the $\theta(\bm r, t)$ is orders of magnitude smaller than that of its static counterpart $\theta_0$ in condensed-matter systmes, making it highly challenging to detect DAQs through their spectroscopic signatures \cite{wu2016quantized, qiu2025observation}.

On the other hand, bound states in the continuum (BICs), first proposed by von Neumann and Wigner in quantum mechanics \cite{neumann1929merkwurdige}, provide a unique mechanism for achieving optical resonances with high quality ($Q$) factors and enhanced light-matter interactions \cite{hsu2016bound,kang2023applications,zeng2024manipulating,wang2024optical}. BICs can arise at the $\Gamma$ point in momentum space protected by $C_{2z}$ symmetry \cite{koshelev2018asymmetric,OvervigPhysRevB}. Breaking this symmetry converts BICs into quasi-BICs (qBICs), enabling a wide range of phenomena, including nonlinear responses, chiral emission, sensing, strong coupling, displacement detection and so on \cite{tittl2018imaging,carletti2018giant,liu2019high,koshelev2019nonlinear,overvig2021chiral,panmai2022highly, zhang2022chiral,weber2023intrinsic,deng2025chiral,zhang2026non}. The breaking of $C_{2z}$ symmetry is typically realized by introducing geometric asymmetry in the structure.

Here, we reveal that DAQ-mediated magnetoelectric coupling can induce an optical shear effect (off-diagonal terms in permittivity tensor) in optical metagratings, enabling a reconfigurable breaking of $C_{2z}$ symmetry via an external static magnetic field without reducing the structural symmetry. Based on this mechanism, we show that DAQs convert symmetry-protected BICs into qBICs, forming axion qBICs that enable a all-optical route for detecting DAQs and significantly enhancing DAQ-mediated second-harmonic generation (SHG). In our approach, we elucidate that the evolution of the $Q$ factor and topological defects in momentum space, together with the emergence of optical resonances in far field associated with axion qBICs, provides new optical signatures for DAQ detection beyond widely adopted Kerr and Faraday effects. Furthermore, we show that axion qBICs can remarkably enhance the SHG conversion efficiency beyond $10^{-6}~W^{-1}$ via critical coupling, providing axion materials with a new platform for exploring nonlinear optical phenomena via magnetoelectric coupling. Our findings establish a symmetry-sensitive BIC mechanism for DAQ detection and demonstrate that axion materials constitute a versatile platform for dynamically controlling light-matter interactions and nonlinear optics.

\begin{figure*}\includegraphics[width=0.85\textwidth]{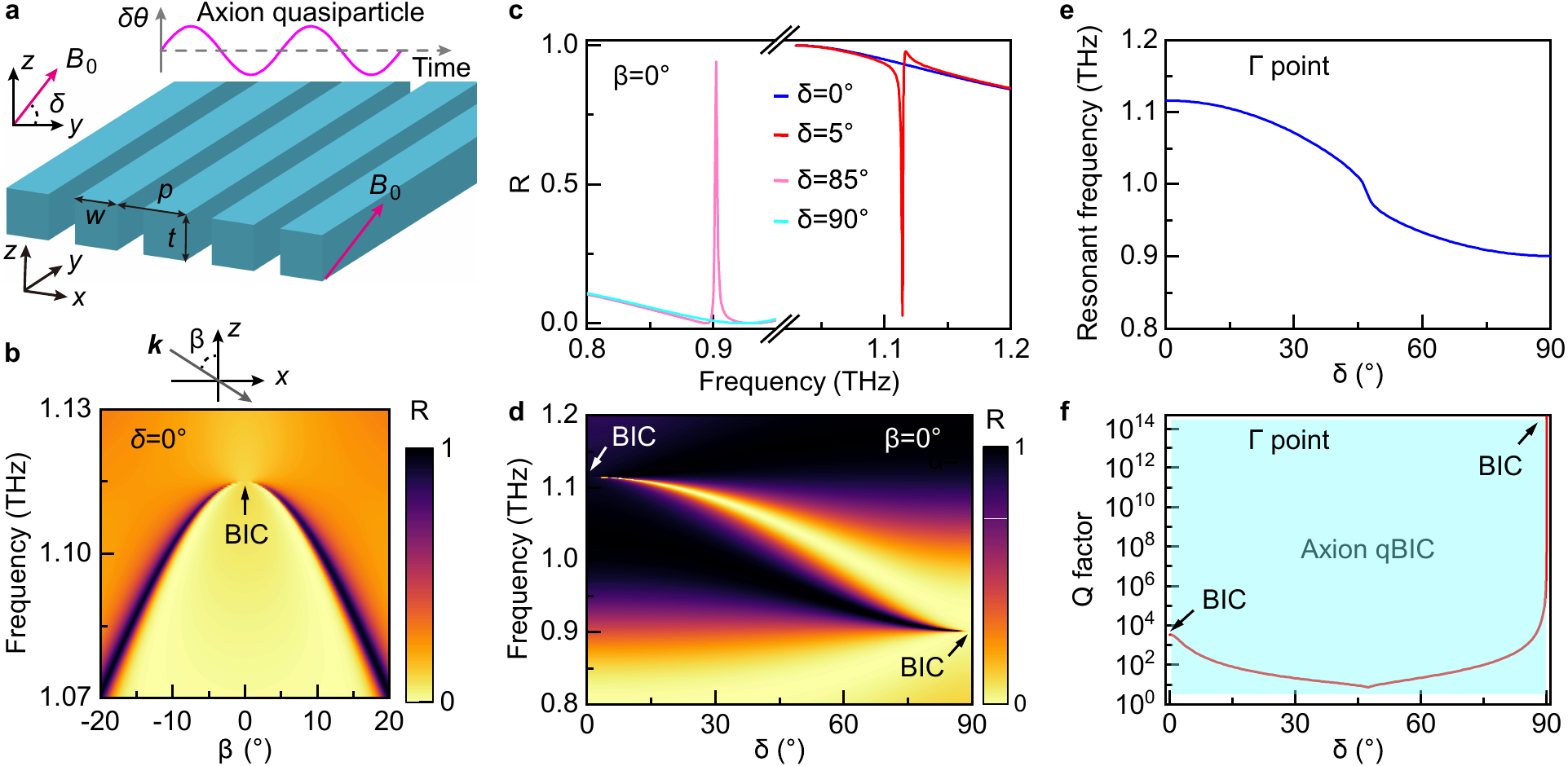}
\caption{Axion quasi-bound states in the continuum. (a) Schematic illustration of the one-dimensional (1D) metagratings hosting dynamic axion quasiparticles (DAQs). $p$ denotes the lattice period, and $t$ and $w$ represent the thickness and width of the axion materials, respectively. $\delta\theta$ denotes the dynamical part of the axion field. The external static magnetic field $\bm B_0$ lies in the $yz$-plane, and $\delta$ represents the angle between $\bm B_0$ and the $y$-axis. (b) Calculated co-polarized angle-resolved reflection spectra for the 1D metagratings in (a) under $y$-polarized incident light. The incident light lies in the $xz$ plane, with $\beta$ representing the angle between the wave vector $\bm k$ and the $z$ axis. (c, d) Co-polarized reflection spectra of normally incident $x$-polarized light versus $\delta$ and the incident light frequency. (e, f) Variation of the axion qBIC frequency in (e) and $Q$ factor in (f) with $\delta$. $p=140~\mu$m, $w=0.6p$ and $t=0.5p$ for the results in (b-f).}
\label{Fig-1}
\end{figure*}

We pattern materials hosting DAQs into one-dimensional (1D) metagratings, as shown in Fig~\ref{Fig-1}(a).  $p$, $t$ and $w$ are the period, thickness and width of the axion material in the 1D metagratings, respectively. The effective action $S$ that characterizes the dynamical axion-photon coupled system is given by,
\begin{equation}
\begin{aligned}
  S&=\frac{1}{8\pi}\int (\varepsilon \bm E^2-\bm B^2/\mu)dx^3dt+\frac{\alpha}{4\pi^2}\int  (\theta_0+\delta\theta)\bm E\cdot\bm Bdx^3dt\\
   &+J\int [(\partial_t\delta\theta)^2-(v_i\partial_i\delta\theta)^2-\Omega_0^2\delta\theta^2]dx^3dt
\end{aligned}
\label{S-action}
\end{equation}
where $\varepsilon$ ($\mu$), $\theta_0$ and $\alpha$ denote the permittivity (permeability), static part of the axion field and fine-structure
constant, respectively. $\delta\theta$ represents the dynamic component of the axion field, varying in space and time, as shown in Fig~\ref{Fig-1}(a). $J$, $\Omega_0$ and $v_i$ are the stiffness, mass and velocity of DAQs, respectively. Minimizing the $S$ in Eq.~\ref{S-action} yields the axion electrodynamics of DAQs via the Euler-Lagrange equation, 
\begin{equation}
\begin{aligned}
  \nabla\cdot\bm D&=-\frac{\alpha}{Z_0\pi}\nabla(\theta_0+\delta\theta)\cdot\bm B\\
  \nabla\times\bm H&=\frac{\partial\bm D}{\partial t}+\frac{\alpha}{Z_0\pi}[\nabla(\theta_0+\delta\theta)\times\bm E+\frac{\partial (\theta_0+\delta\theta)}{\partial t} \bm B]\\
  \nabla\times\bm E&=-\frac{\partial\bm B}{\partial t}\\
  \nabla\cdot\bm B&=0\\
  \kappa\bm E\cdot\bm B&=\frac{\partial^2\delta\theta}{\partial t^2}-v^2\nabla^2\delta\theta+\Omega_0^2\delta\theta
\end{aligned}
\label{AE}
\end{equation}

We have $\kappa=\alpha/8\pi^2J$ in SI unit and $Z_0$ is the characteristic impedance of vacuum. The total magnetic field $\bm B=\bm B_0+\bm B (\bm r, t)$ comprises a static, uniform component $\bm B_0$ and a dynamical part $\bm B (\bm r, t)$, with $\bm B (\bm r, t)\ll \bm B_0$. We further consider $\theta_0=0$ and that the axion velocity is much smaller than the speed of light. With a time-harmonic dependence $e^{i\omega t}$,  the linearized axion electrodynamics in the frequency domain reads (see Sec.\uppercase\expandafter{\romannumeral2} in the SI),
\begin{equation}
\begin{aligned}
  \nabla\times\nabla\times\bm E_{\omega}=\omega^2\varepsilon\mu[\mathbf{I}-\frac{\kappa\alpha\bm B_0\bm B_0^T}{(\omega^2-\Omega_0^2-i\omega\gamma_a)(Z_0\pi\varepsilon)}]\bm E_{\omega}
\end{aligned}
\label{LAE-W}
\end{equation}
$\gamma_a$ is phenomenologically introduced to describe the damping of DAQs, which is related to the magnon lifetime \cite{Bayrakci2013}. $\mathbf{I}$ is the identity matrix. We obtain $\bm B_0\bm B_0^T$ in the following form
\begin{equation}
\bm B_0\bm B_0^T=\left(
  \begin{array}{cccc}
    B_0^xB_0^x & B_0^xB_0^y &B_0^xB_0^z\\
    B_0^yB_0^x & B_0^yB_0^y &B_0^yB_0^z \\
    B_0^zB_0^x & B_0^zB_0^y &B_0^zB_0^z\\
  \end{array}
\right).
\end{equation}
where $B_0^i$ is the $i$-component of the magnetic field $\bm B_0$ ($i=x,y,z$). Throughout this work, we take $\varepsilon=10\varepsilon_0$ and $\mu=\mu_0$ for simplicity. 

 We define an effective permittivity tensor $\varepsilon_{eff}=\varepsilon[\mathbf{I}-\kappa\alpha\bm B_0\bm B_0^T/(\omega^2-\Omega_0^2-i\omega\gamma_a)(Z_0\pi\varepsilon)]$ arising from fluctuations of magnetic order. When $\bm B_0$ is not aligned with the principal axes of the 1D metagrating, $\varepsilon_{eff}$ exhibits nonzero off-diagonal components, $\varepsilon_{{eff}}^{ij}\neq 0$ ($i\neq j$). The off-diagonal components $\varepsilon_{eff}^{ij}$ give rise to a $i$-directed polarization perpendicular to the $j$-directed electric field, analogous to shear phenomena in mechanics and viscous flow \cite{sofronis1995mechanics,wu2020energy}. Therefore, $\varepsilon_{eff}^{ij}$ ($i\neq j$) represents an optical shear \cite{ji2026janus,passler2022hyperbolic,hu2023real,renzi2024hyperbolic}. This establishes DAQ-induced magnetoelectric coupling as a new physical mechanism for generating optical shear.. Here, we investigate how such shear effects can be exploited to control axion-photon coupling, providing a new all-optical route for detecting DAQs. Since the magnetic field $\bm B$ is a pseudovector, it exhibits distinct symmetry transformations under mirror operations. Under the mirror operations $\sigma_x$ and $\sigma_y$, we obtain $\sigma_x \bm B_0=(B_0^x,-B_0^y,-B_0^z)^T$ and $\sigma_y \bm B_0=(-B_0^x,B_0^y,-B_0^z)^T$ ($T$ denotes the transpose operation). Consequently, the $C_{2z}=\sigma_x\sigma_y$ symmetry is effectively broken by the $B_0^xB_0^z$ and $B_0^yB_0^z$ contributions in $\varepsilon_{eff}$, even though the geometric structure itself remains symmetric. Thus, the $C_{2z}$ symmetry breaking arising from axion-photon coupling preserves the geometric symmetry and allows for dynamically reconfigurable structural symmetry [Fig~\ref{Fig-1}(a)], distinguishing it from prior schemes relying on mirror-asymmetric geometries \cite{koshelev2018asymmetric,tittl2018imaging,carletti2018giant,liu2019high,koshelev2019nonlinear,overvig2021chiral,zhang2022chiral,weber2023intrinsic,deng2025chiral,zhang2026non}. Notably, the static axion field $\theta_0$ does not break the $C_{2z}$ symmetry, and is therefore neglected in our analysis.

\begin{figure}\includegraphics[width=0.45\textwidth]{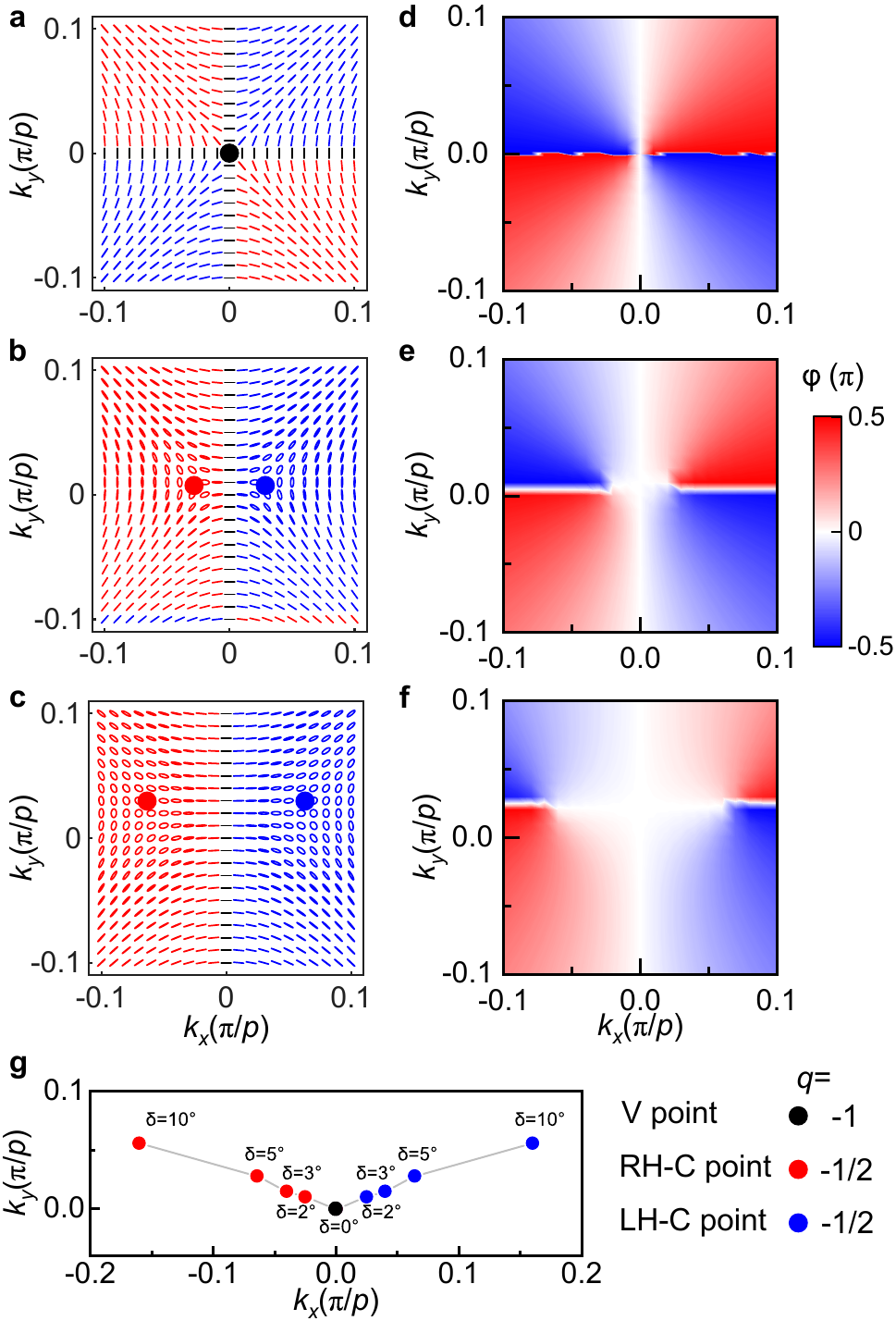}
\caption{Probing DAQs via topological defects in momentum space. (a-c) Simulated polarization state distributions in momentum space. (d-f) Simulated Stokes phase distributions in momentum space. The values of $\delta$ are equal to $0\degree$, $2\degree$ and $5\degree$ for (a, d), (b, e) and (c, f), respectively. Other parameters are identical to those in Fig.1. (g) Momentum-space topological defects versus $\delta$.}
\label{Fig-2}
\end{figure}

We first consider the case where $\bm B_0$ lies in the $yz$ plane, i.e., $\bm B_0=B(0,\cos \delta,\sin \delta)^T$, as shown in Fig~\ref{Fig-1}(a). Here, we define $b=\sqrt{\kappa\alpha B^2/Z_0\pi\varepsilon|}$, characterizing the axion-photon coupling strength, which can be adjusted from 0 to 5~meV/$\hbar$ for materials hosting DAQs \cite{xiao2021nonlinear,zhu2022axionic}. Here, we take $b=2.5$~meV/$\hbar$, $\Omega_0=2$~meV/$\hbar$ and $\gamma_a=0.002$~meV/$\hbar$. Fig~\ref{Fig-1}(b) presents the co-polarized angle-resolved reflection spectra for $y$-polarized incident light, with $\delta=0\degree$ and the incident plane lying in the $xz$ plane. In such case, the 1D metagrating retains $C_{2z}$ symmetry in Fig~\ref{Fig-1}(a). The resonance linewidth increases with incident angle $|\beta|$, while it vanishes at $\beta=0\degree$, revealing a $C_{2z}$ symmetry-protected BIC at the $\Gamma$ point [see Fig~\ref{Fig-1}(b)]. 

Next, we examine the reflection spectra of BICs for normally incident $x$-polarized light vary with $\delta$ [Fig.~\ref{Fig-1}(c, d)]. It can be observed that for $\delta=5\degree$ and $\delta=85\degree$, the reflection spectra exhibit pronounced Fano-like lineshapes with narrow spectral linewidths, whereas for $\delta=0\degree$ and $\delta=90\degree$, the Fano-like lineshapes of reflection spectra vanish, as shown in Fig.~\ref{Fig-1}(c). These results demonstrate that DAQ-induced optical shear can transform BICs into qBICs, which we term axion qBICs. Fig~\ref{Fig-1}(d) illustrates the detailed evolution of the reflection spectrum of the axion qBICs as a function of $\delta$. The reflection dip induced by the axion qBIC gradually evolves into a reflection peak as 
$\delta$ increases [Fig.~\ref{Fig-1}(c, d)]. In addition, the resonance linewidth first broadens and then narrows with increasing $\delta$, eventually vanishing at $\delta=90\degree$, as shown in Fig~\ref{Fig-1}(d). The resonance frequency of the axion qBIC decreases with increasing $\delta$, as shown in Fig~\ref{Fig-1}(e). 

We further extract the $Q=\omega_0/2(\gamma_r+\gamma_{nr})$ factor of the axion qBIC, as shown in Fig.~\ref{Fig-1}(f). Here, $\omega_0$, $\gamma_r$ and $\gamma_{nr}$ denote the resonance angular frequency, radiative loss rate and nonradiative loss rate of the axion qBIC, respectively. The $\gamma_r$ is determined by the $C_{2z}$ symmetry breaking induced by DAQs, while the $\gamma_{nr}$ is related to axion scattering $\gamma_{a}$. A BIC is characterized by zero radiative loss, i.e., $\gamma_r=0$. As shown in Fig.~\ref{Fig-1}(f), the $Q$ factor is highly sensitive to the $C_{2z}$ symmetry breaking induced by DAQs. At $\delta=0\degree$ and $\delta=90\degree$, the $\Gamma$-point modes of the 1D metagrating are either transverse electric (TE) with $\bm E_{\omega}=(0,E_y,0)$ or transverse magnetic (TM) with $\bm E_{\omega}=(E_x,0,E_z)$ modes. For the TE mode, it only experiences the $\varepsilon_{eff}^{yy}$ component of $\varepsilon_{eff}$. At $\delta=0\degree$, the BIC has a finite $Q$ for the TE mode due to loss from $\varepsilon_{eff}^{yy}$, while at $\delta=90\degree$, $Q$ factor diverges because $\varepsilon_{eff}^{yy}$ is lossless, as shown in Fig.~\ref{Fig-1}(f). Consequently, the $Q$ factor of axion qBICs is highly sensitive to $\delta$ near $90\degree$. Thus, the $Q$ factor and mode emergence in far field provide highly sensitive optical probes for DAQ detection. However, conventional approaches based on chiroptical effects, such as the Faraday and Kerr rotations \cite{wu2016quantized,okada2016terahertz,ahn2022theory,qiu2025observation}, require high-precision terahertz polarimetry, as the resulting rotation angles, constrained by the fine-structure constant, are exceedingly small and lie below the sensitivity of conventional terahertz measurements, posing significant experimental challenges. Our approach exploits non-chiroptical effects associated with axion qBICs, thereby effectively circumventing these experimental challenges.

\begin{figure*}\includegraphics[width=0.85\textwidth]{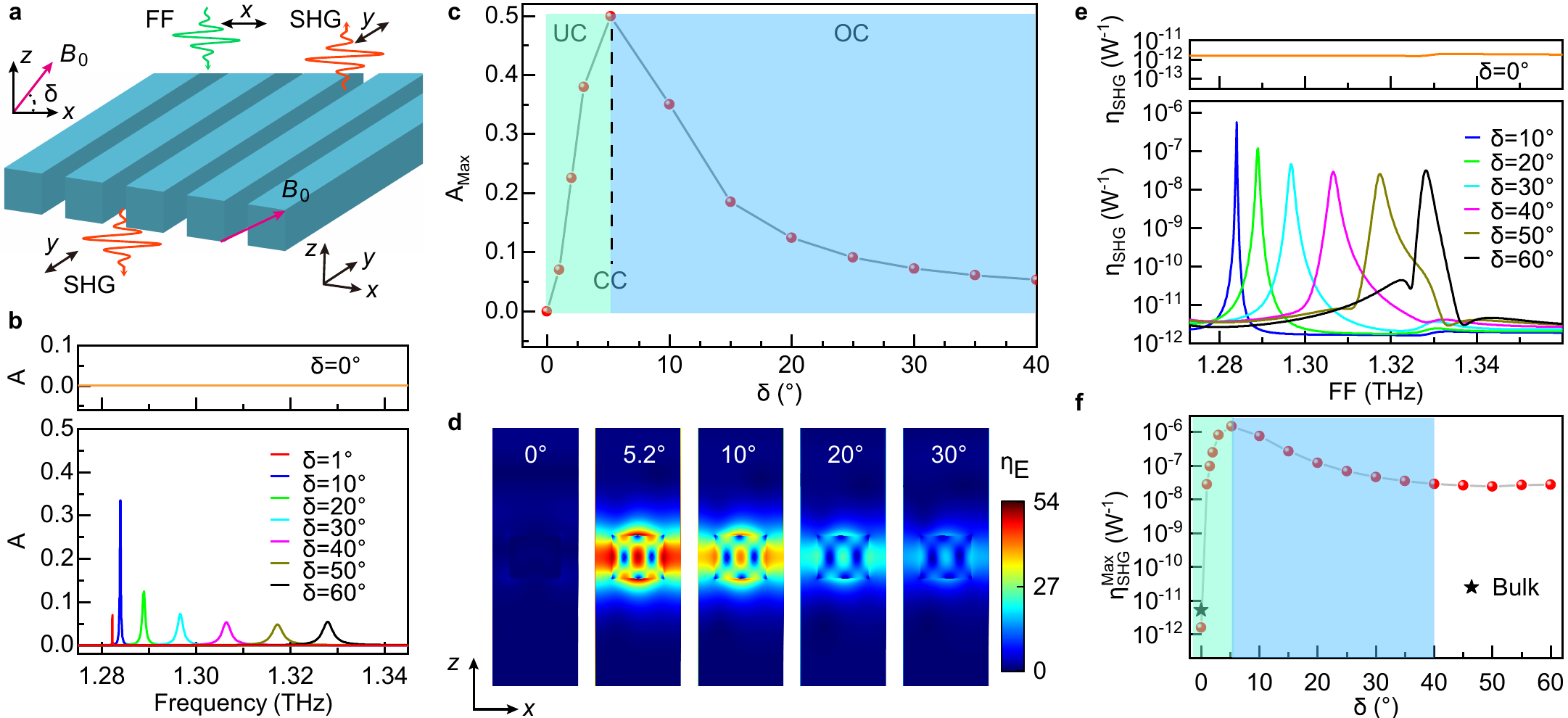}
\caption{DAQ-mediated second-harmonic generation (SHG). (a) Schematic illustration of the SHG in 1D metagratings hosting DAQs. $\bm B_0$ lies in the $xz$ plane and makes an angle of $\delta$ with the $x$ axis. The fundmental frequency (FF) wave is polarized along the $x$ direction, while the SHG is polarized along the $y$ direction for TM mode. (b) Absorption spectra $A$ versus $\delta$ under normally incident $x$-polarized light for TM mode. (c) Absorption peak $A_{max}$ versus $\delta$. (d) Electric-field enhancement $\eta_E$ versus $\delta$. (e) SHG efficiency $\eta_{\mathrm{SHG}}$ verus $\delta$. (f) Maximum SHG efficiency$\eta_{\mathrm{SHG}}^{\mathrm{Max}}$ verus $\delta$. The structural parameters are identical to those presented in Fig.1. $B=2$ T.}
\label{Fig-3}
\end{figure*}

We further reveal that topological defects in momentum space can serve as another optical signature for detecting DAQs, as shown in Fig.~\ref{Fig-2}. BICs and circular polarization states (C points) are widely recognized as topological defects in momentum space \cite{zhen2014topological,liu2019circularly,ji2024probing}, carrying a topological charge $q=1/2\pi \oint_{L} d\bm k_{\varparallel}\cdot \nabla_{\bm k_{\varparallel}}\varphi(\bm k_{\varparallel})$. $\varphi(\bm k_{\varparallel})$ is the Stokes phase, defined as the angle bewteen the long axis of polarization ellipse and the $k_x$ axis. Fig.~\ref{Fig-2}(a) shows that the polarization states around the BIC form a saddle-like vortex pattern, with the BIC located at the vortex center (V point). From the vortex configuration and the Stokes phase $\varphi(\bm k_{\varparallel})$ distribution in Fig.~\ref{Fig-2}(a, c), the BIC is identified to carry a topological charge of $q=-1$. At $\delta=2\degree$, the breaking of $C_{2z}$ symmetry causes the V point to split into a pair of left- and right-handed (LH and RH) C points, each carrying a topological charge of $q=-1/2$, as shown in Fig.~\ref{Fig-2}(b, d). As $\delta$ increases to $5\degree$, the pair of C points moves away from the $\Gamma$ point toward larger wave vectors. Owing to topological robustness, these C points persist and do not disappear as $\delta$ increases, as shown in Fig.~\ref{Fig-2}(g). Thus, symmetry-sensitive V points (BICs) and robust C points serve as highly sensitive and reliable optical probes for DAQ detection.

Finally, we investigate SHG enhanced by axion qBICs. The nonlinear current $\bm J_{NL}^{2\omega}$ induced by DAQs is given by (see Sec.\uppercase\expandafter{\romannumeral3} in the SI),
\begin{equation}
\begin{aligned}
  \bm J_{NL}^{2\omega}&=-\nabla\times(\Lambda (\omega)\bm B_{0}\cdot\bm E_{\omega}\bm E_{\omega})-2i\omega\Lambda (\omega)\bm B_{0}\cdot\bm E_{\omega}\bm B_{\omega}\\
  &-2i\omega\Lambda (2\omega)\bm B_{\omega}\cdot\bm E_{\omega}\bm B_{0}
\end{aligned}
\label{NAE-W}
\end{equation}
where $\Lambda (\omega)=\kappa\alpha/Z_0\pi(\omega^2-\Omega_0^2-i\omega\gamma_a)$.
As seen from Eq.~\ref{NAE-W}, DAQ-induced SHG consists of three contributions, originating from the nonlinear magnetization $\bm M_{NL}^{2\omega}=\alpha\delta\theta_{\omega}\bm E_{\omega}/Z_0\pi$, the nonlinear polarization $\bm P_{NL}^{2\omega}=\alpha\delta\theta_{\omega}\bm B_{\omega}/Z_0\pi$, and the axionic nonlinearity $\delta\theta_{NL}^{2\omega}=\Lambda (2\omega)\bm B_{\omega}\cdot\bm E_{\omega}$, respectively. Therefore, the DAQ-induced SHG originates from the magnetoelectric coupling mediated by the dynamical axion field and the conversion of light into the dynamical axion field via two-photon process. This is in contrast to conventional nonlinear materials, where SHG is governed solely by the electric-field response in nonlinear polarization $\bm P_{NL}^{2\omega}=\bm{\chi}^{(2)}:\bm E_{\omega}\bm E_{\omega}$ \cite{vabishchevich2023nonlinear}.

Here, we consider the TM mode in the 1D metagratings and $\bm B_0$ confined to the $xz$ plane, $\bm B_0=B(\cos \delta, 0,\sin \delta)^T$, as shown in Fig~\ref{Fig-3}(a). In this case, $\bm J_{NL}^{2\omega}$ is along $y$ direction for an $x$-polarized fundamental pump. Therefore, the polarization of the second-harmonic mode is completely orthogonal to that of the fundamental light, enabling DAQs to be probed via SHG polarization.

At $\delta=0\degree$, absorption $A$ is weak, whereas at $\delta=1\degree$, a sharp absorption peak appears in the 1D metagrating, as shown in Fig~\ref{Fig-3}(b). This reveals a TM-mode BIC that can be transformed into an axion qBIC through DAQ-induced coupling. According to coupled-mode theory, the absorption spectrum  $A$ can be expressed as, $A=2\gamma_r\gamma_{nr}/[(\omega-\omega_0)^2+(\gamma_r+\gamma_{nr})^2]$ \cite{hong2023nonlocal}. The $\gamma_r$ and $\omega_0$ can both be controlled by $\delta$, as shown in Fig~\ref{Fig-3}(b). Therefore, by tuning the ratio between $\gamma_r$ and $\gamma_{nr}$ of the axion qBICs, the system can be driven across the under-coupled (UC) with $\gamma_r<\gamma_{nr}$, critically coupled (CC) with $\gamma_r=\gamma_{nr}$, and over-coupled (OC) with $\gamma_r>\gamma_{nr}$ regimes, as evidenced by the maximum absorption $A_{max}$ in Fig~\ref{Fig-3}(c). Such continuous control of the coupling regime allows tuning of the fundamental-wave electric-field enhancement, $\eta_E^2=|\bm E_{\omega}|^2/|\bm E_{\omega}^{inc}|^2\propto \gamma_r/(\gamma_r+\gamma_{nr})^2$, where $|\bm E_{\omega}|$ is the amplitude under excitation $\bm E_{\omega}^{inc}$ \cite{liang2020bound}. As shown in Fig~\ref{Fig-3}(d), the $\eta_E$ under CC condition ($\delta=5.2\degree$) is significantly greater than in the UC and OC regimes, thereby tuning the DAQ-induced SHG response. We define the normalized SHG conversion efficiency as $\eta_{\mathrm{SHG}}=P_{\mathrm{SHG}}/P_{\mathrm{pump}}^2$, where $P_{\mathrm{SHG}}$ and $P_{\mathrm{pump}}$ denote the generated SHG power and the pump power, respectively. The results for $\eta_{\mathrm{SHG}}$ and the maximum SHG efficiency $\eta_{\mathrm{SHG}}^{\mathrm{Max}}$ are shown in Fig~\ref{Fig-3}(e) and Fig~\ref{Fig-3}(f), respectively. It can be clearly seen that the SHG efficiency reaches its maximum under the CC condition, exceeding $10^{-6}~W^{-1}$, despite the 1D metagrating being only about one-third of the operating wavelength in thickness. However, for axionic materials with the same thickness, the SHG efficiency is only on the order of $10^{-12}~W^{-1}$, as indicated by the black star in Fig~\ref{Fig-3}(f). Thus, the incorporation of BIC physics offers a promising strategy to significantly enhance axion-induced SHG.

In summary, chiral manifestations of DAQs in condensed matter systems, such as Kerr and Faraday rotations and polarization conversion, are intrinsically weak due to an extremely small dynamic axion field \cite{qiu2025observation,liebman2025multiphoton}. These fundamental limitations have thus far hindered the direct optical probing of DAQs and practical utilization of optoelectronic devices. Here, we elucidate DAQ-induced optical shear and propose axion qBICs for all-optical definitive DAQ detection based on achiral manifestations. In our scheme, the quality factor, mode emergence and topology in far field provide sensitive achiral optical signatures for DAQs probing, distinct from existing approaches. In addition, we show that materials hosting DAQs are ideal candidates for exploring nonlinear optics via magnetoelectric coupling, governed by a physical mechanism distinct from that of conventional nonlinear materials such as GaAs, BaTiO$_3$, LiNbO$_3$, MoS$_2$, a-Se and others \cite{vabishchevich2023nonlinear}. It should be noted that the reported phenomena do not rely on specific material parameters or material platforms, suggesting their general applicability across a broad class of magnetoelectric systems.

Axion dark matter searches in the meV mass range are particularly promising, as motivated by cosmology and astrophysics \cite{gorghetto2021more,saikawa2024spectrum}. However, viable detectors for this regime remain lacking. In this context, the proposed axion qBIC physics may serve as a high-Q, sensitive resonant detector for cosmic axion searches in the meV range.

Dynamically reconfigurable symmetry breaking is essential for advanced optoelectronics applications. However, conventional phase-change and electro-optic materials remain inadequate for achieving dynamic symmetry control in photonic systems \cite{zhang2021electrically,xie2024ultrathin,soma2025subvolt}. Our results also uncover that axion materials provide a new magnetoelectric coupling mechanism for dynamic symmetry control, opening up opportunities for reconfigurable tuning of topology, vortex generation, polarization and chirality \cite{wang2020generating,ji2026intrinsically}.

{\textbf{Acknowledgments:}  The authors acknowledge the support from the National Natural Science Foundation of China (Grant Nos. W2511003, 12321004, 12204041, 62375016, T2325005, 12274030, 52488301, 12574335, 12234003)}, the Fundamental and Interdisciplinary Disciplines Breakthrough Plan of the Ministry of Education of China (Grant No. JYB2025XDXM115) and National Key Research and Development Program of China (Grant No. 2022YFA1403400, 2024YFB2809204).


\end{document}